\documentclass[twocolumn,superscriptaddress,nofootinbib]{revtex4-2}

\usepackage{amsmath,amssymb}
\usepackage{graphicx}
\usepackage{hyperref}
\usepackage{xcolor}

\usepackage[final]{paperclaims}

\newcommand{\repourl}{https://github.com/matiaszaldarriaga/pta-gwb-anisotropy}
\newcommand{\zenododoi}{10.5281/zenodo.21911100}
\newcommand{\zenodourl}{https://doi.org/\zenododoi}

\begin{document}

\title{Anisotropies in the PTA gravitational wave background: what can they teach us about supermassive black hole binaries?}

\author{Matias Zaldarriaga}
\affiliation{School of Natural Sciences, Institute for Advanced Study, Princeton, NJ 08540, USA}
\author{Gabriela Sato-Polito}
\affiliation{School of Natural Sciences, Institute for Advanced Study, Princeton, NJ 08540, USA}

\date{\today}

\begin{abstract}
The gravitational wave background detected by pulsar timing arrays is
sourced by a finite population of supermassive black hole binaries, and is
therefore anisotropic. We ask what measuring that anisotropy can teach us
about the population, using models that span a wide range of effective
source counts, all normalized to the measured background amplitude. We find
four things. First, the expected anisotropy is produced by the single
brightest binary: a dipole at the level of the published NANOGrav $95\%$
upper limit would require one source to supply about $60\%$ of the power in
the band. Second, no model that also reproduces the measured strain spectrum
contains a source that bright. In every case the loudest binary stays below
the NANOGrav continuous-wave upper limit at every frequency it covers,
consistent with the joint search of the 15-year data, which finds no
resolved source. Third, because the anisotropy is produced by one source,
compressing the sky to an angular power spectrum discards the phase
information that locates it. Such a search is never more sensitive than
looking for the source directly, and is strictly worse once more than a
dipole is kept. Fourth, the published upper limits on the angular power
spectrum therefore reflect the analysis prior rather than the data: they
coincide with the 95th percentile of the prior induced by the square-root
spherical harmonic basis adopted in the analysis. What does constrain the
population today is the shape of the strain spectrum. Rare, bright sources
depress the median spectrum below its mean, and fitting the measured
spectrum already disfavors mass functions dominated by binaries above about
$10^{10}\,M_\odot$.
\end{abstract}

\maketitle

\section{Introduction}
\label{sec:intro}

Pulsar timing arrays (PTAs) have reported strong evidence for a
nanohertz gravitational wave background
(GWB)~\cite{NANOGrav:2023gor,EPTA:2023fyk,Reardon:2023gzh,Xu:2023wog}.
The signal is consistent with expectations from a cosmic population of
inspiraling supermassive black hole binaries (SMBHBs), and its spectral
properties are beginning to constrain the demographics of SMBH
populations~\cite{SatoPolito:2023big,SatoPolito:2025dist,Liepold:2024big}.

A natural next step is to search for anisotropies in the GWB. A finite
population of discrete sources at specific sky locations produces
angular structure in the gravitational wave power, characterized by
angular power spectra $C_\ell$. Several groups have proposed using
$C_\ell$ measurements to constrain SMBHB population
models~\cite{Mingarelli:2013dsa,Taylor:2013esa,SatoPolito:2024Kam}, and
the NANOGrav collaboration has published upper limits on $C_\ell/C_0$
from their 15-year dataset~\cite{NANOGrav:2023tcn}. Recent work has
compared the expected shot-noise anisotropy of empirically calibrated
merger models with those bounds~\cite{LinLidzMa:2026}, and with its
realization-to-realization distribution~\cite{LinLidzMa:2026b}.

In this paper we ask what anisotropy measurements can teach us about
the astrophysical model, given the data we have and the data coming
soon.\footnote{All of the code, data and \LaTeX{} source behind this
paper are publicly available at \url{\repourl}, and the production
Monte Carlo arrays are archived at \url{\zenodourl}. The repository
carries detailed provenance information: every number quoted in the
text resolves to the code that computed it, every figure to the
script and the input arrays that produced it, and every data product
records the exact command line, random seed and checksum of the run
that made it, together with a single command that rebuilds the whole
analysis from those seeds.} We find that, for the astrophysically
expected signal,
essentially all of the anisotropy is produced by the single brightest
source. There is an exact relation between the angular power spectrum
of a sky made of discrete sources and the fractions of the total power
that individual sources carry, and it can be read in either direction.
A measured dipole implies a definite brightest-source fraction and a
definite effective number of sources, almost independently of the
population model. A dipole at the level of the published NANOGrav
$95\%$ value would require a single binary carrying about $60\%$ of the
power in the band (Sec.~\ref{sec:dipole_source}).

In every astrophysical model we consider, including the most
anisotropic ones, that brightest binary is too faint to be detected at
current sensitivity. Its strain stays below the sky-averaged NANOGrav
continuous-wave upper limit at every frequency that limit covers, and a
joint search for
a resolved source in the 15-year data finds
none~\cite{Goncharov:2026joint}. The best way to detect the anisotropy
is to search for the source itself: the optimal search has significance
$\rho_{\rm ps} = \sqrt{5}\,p_1\,\rho_0$, set by the brightest-source
fraction $p_1$ and the signal-to-noise ratio $\rho_0$ of the
Hellings--Downs cross-correlation. For the NANOGrav 15-year data $\rho_0 \approx 5$
integrated over the entire band~\cite{NANOGrav:2023gor}, and
$\rho_0 \approx 2$ in each of the roughly five bins that carry the
signal, which explains why no source is close to being detected.
Compressing the anisotropy to an angular power spectrum is never more
sensitive (equal for a dipole-only test, weaker once additional modes
are included), because it discards the phase information that locates
the source. If the optimal search cannot find the source, no $C_\ell$
statistic can find its anisotropy (Sec.~\ref{sec:source_detection}).

This also explains where the published $C_\ell$ upper limits come from.
Since the data cannot at present inform any anisotropy statistic, a
Bayesian upper limit can only return the prior. We show that the
NANOGrav bound $C_1/C_0 \simeq 0.2$ coincides with the 95th percentile
of the prior induced by the square-root spherical harmonic basis used
in the analysis, and that this prior depends sensitively on an
arbitrary truncation parameter (Sec.~\ref{sec:nanograv}).

We conclude that anisotropy does not constrain the SMBHB population
today, and will not until the background cross-correlation is measured
far more significantly. What already constrains the population is the
shape of the strain spectrum: the rare, bright sources that would make
the background anisotropic also depress the median realization below
the $f^{-4/3}$ ensemble mean by a model-dependent amount that grows
with frequency. Fitting the NANOGrav 15-year free spectrum to the
predicted strain distribution already requires roughly ten times more
black holes than local estimates suggest and disfavors mass functions
dominated by very heavy binaries, $M_{\rm peak}\gtrsim
10^{10}\,M_\odot$~\cite{SatoPolito:2025dist}, with no anisotropy
measurement involved.

The paper is organized as follows. Section~\ref{sec:cl} collects exact
formulas for the angular power spectrum of a discrete-source sky.
Section~\ref{sec:astro} describes the astrophysical population model,
the Monte Carlo realizations, and the frequency grid and observation
time we adopt. Section~\ref{sec:dipole_source} shows that a large
dipole and a single dominant source are equivalent statements, and that
the dominant source is individually faint at present.
Section~\ref{sec:source_detection} derives the significance of the
optimal source search, anchors it to the measured background
signal-to-noise, and compares the $C_\ell$ compression.
Section~\ref{sec:nanograv} compares the published $C_\ell$ upper limits
with the prior of the analysis basis. Section~\ref{sec:discussion}
summarizes the resulting observational strategy.
\section{Angular power spectrum from discrete sources}
\label{sec:cl}

\subsection{Source-level formulas}

The angular power spectrum of a sky made of point sources can be written
down exactly. No expansion in the number of sources or in their
brightness contrast is needed, and the formulas collected here are the
ones used throughout the paper.

Consider a sky map of GW power from $N_s$ discrete sources,
\begin{equation}
M(\hat{n}) = \sum_{a=1}^{N_s} q_a \,\delta^{(2)}(\hat{n}, \hat{n}_a),
\end{equation}
where $q_a > 0$ is the weight (e.g., $h^2_s$) of source $a$ at sky
direction $\hat{n}_a$. The spherical harmonic coefficients are
\begin{equation}
a_{\ell m} = \sum_a q_a \, Y_{\ell m}^*(\hat{n}_a),
\end{equation}
and the angular power spectrum is
\begin{equation}
C_\ell = \frac{1}{2\ell+1} \sum_{m=-\ell}^{\ell} |a_{\ell m}|^2.
\end{equation}
Using the addition theorem, the exact source-level formula is
\begin{equation}
\label{eq:cl_exact}
C_\ell = \frac{1}{4\pi} \sum_{a,b} q_a q_b \, P_\ell(\hat{n}_a \cdot \hat{n}_b),
\end{equation}
where $P_\ell$ is the Legendre polynomial. In particular,
\begin{equation}
C_0 = \frac{1}{4\pi}\left(\sum_a q_a\right)^2.
\end{equation}

\subsection{Normalized weight fractions}

Define the total source weight $Q \equiv \sum_a q_a$ and the normalized
fractions
\begin{equation}
p_a \equiv \frac{q_a}{Q}, \qquad \sum_a p_a = 1.
\end{equation}
Then
\begin{equation}
\label{eq:cl_c0_exact}
\frac{C_\ell}{C_0} = \sum_{a,b} p_a p_b \, P_\ell(\hat{n}_a \cdot \hat{n}_b).
\end{equation}
This expression is exact: the ratio $C_\ell/C_0$ depends on the source
amplitudes only through the normalized fractions $p_a$, not through the
overall scale $Q$.

\subsection{The effective number of sources}

Define the effective number of sources as the inverse participation ratio,
\begin{equation}
\label{eq:neff}
N_{\rm eff} \equiv \frac{1}{\sum_a p_a^2}.
\end{equation}
For $N$ identical sources, $N_{\rm eff} = N$. When one source dominates,
$N_{\rm eff} \to 1$.

The conditional mean (averaging over isotropic source positions at fixed
weights) is
\begin{equation}
\label{eq:mean_cl}
\mathbb{E}\!\left[\frac{C_\ell}{C_0} \;\middle|\; \{p_a\}\right]
= \sum_a p_a^2 = \frac{1}{N_{\rm eff}}
\qquad (\ell \geq 1),
\end{equation}
since $\mathbb{E}[P_\ell(\hat{n}_a \cdot \hat{n}_b)] = 0$ for
$a \neq b$ and $P_\ell(1) = 1$. The mean is the same for all
$\ell \geq 1$.

\subsection{Specialization to the dipole}

For $\ell = 1$, $P_1(\mu) = \mu$, and Eq.~\ref{eq:cl_c0_exact}
simplifies to
\begin{equation}
\label{eq:dipole}
\frac{C_1}{C_0} = \left|\sum_a p_a \hat{n}_a\right|^2 \equiv |\mathbf{S}|^2,
\end{equation}
where $\mathbf{S} \equiv \sum_a p_a \hat{n}_a$ is a weighted random walk
in $\mathbb{R}^3$ with step lengths $p_a$ and random isotropic
directions. The conditional variance is
\begin{equation}
\mathrm{Var}\!\left[\frac{C_1}{C_0} \;\middle|\; \{p_a\}\right]
= \frac{2}{3}\left[\left(\sum_a p_a^2\right)^2 - \sum_a p_a^4\right].
\end{equation}
The distribution of $\mathbf{S}$ is most easily obtained from its Fourier
transform, $\tilde P(\mathbf{k}) = \langle e^{i\mathbf{k}\cdot\mathbf{S}}
\rangle$. Because the steps of the walk are independent, this is the
product of the transforms of the individual steps. A single step has
fixed length $p_a$ and an isotropic random direction, so its transform is
the direction average $\langle e^{i\mathbf{k}\cdot p_a\hat n}\rangle =
\sin(p_ak)/(p_ak)$, which depends only on $k=|\mathbf{k}|$. Hence
\begin{equation}
\tilde P(k) = \prod_a \frac{\sin(p_a k)}{p_a k},
\end{equation}
and the distribution of $C_1/C_0 = |\mathbf{S}|^2$ follows from a
one-dimensional inverse transform (see Sec.~\ref{sec:dipole_source}).
\section{The astrophysical model}
\label{sec:astro}

\subsection{Population model}

We adopt the SMBHB population model of Sato-Polito and
Zaldarriaga~\cite{SatoPolito:2023big}, which computes the expected
number of sources $\bar{N}(f, M)$ and the strain per source $h^2_s(f, M)$
in bins of GW frequency $f$ and SMBH total mass $M$.

The model starts from the velocity dispersion function of galaxies,
converts to SMBH masses via the $M$--$\sigma$ relation,
$\log_{10}(M/M_\odot) = a + b\,\log_{10}(\sigma/\sigma_{\rm ref})$,
with intrinsic scatter $\varepsilon$ (in dex) around the mean
relation~\cite{SatoPolito:2023big}, and integrates over redshift and
mass ratio to obtain the luminosity function. The key parameter
controlling the high-mass tail is the scatter $\varepsilon$: larger
values populate more massive binaries and produce a louder, more
anisotropic background.

We consider three values:
\begin{itemize}
\item $\varepsilon = 0.20$: $\log_{10}(M_{\rm peak}/M_\odot) = 9.2$
\item $\varepsilon = 0.38$: $\log_{10}(M_{\rm peak}/M_\odot) = 9.5$
  (the value used in~\cite{SatoPolito:2023big})
\item $\varepsilon = 0.66$: $\log_{10}(M_{\rm peak}/M_\odot) = 10.6$
\end{itemize}
These span the range considered in~\cite{SatoPolito:2025dist}.

These three values of $\varepsilon$ illustrate different regimes of
$N_{\rm eff}$ and $p_1$. We normalize all three to the same observed
mean amplitude, $\sqrt{\langle h_c^2\rangle} = 2.5\times10^{-15}$
at $f = 1\,{\rm yr}^{-1}$, so that they differ only in how that fixed
power is distributed among sources. Because the mean total strain scales
linearly with the velocity-dispersion-function normalization $\phi_\sigma$
and exponentially with the scatter, $\langle h_c^2\rangle \propto
\phi_\sigma\,\exp(25\varepsilon^2\ln^2 10/18)$, matching the amplitude
fixes $\phi_\sigma$ for each model. Relative to the fiducial value of
Ref.~\cite{SatoPolito:2023big} this requires multiplying $\phi_\sigma$ by
$\approx 13$, $6$, and $0.7$ for $\varepsilon = 0.20$, $0.38$, and $0.66$.%
\dataref{phi_sigma_factor_eps020}{13.256}%
\dataref{phi_sigma_factor_eps038}{6.145}%
\dataref{phi_sigma_factor_eps066}{0.720}
The low-scatter models thus need a local SMBH abundance well above the
observed one to reach the GWB amplitude, whereas $\varepsilon = 0.66$
slightly overshoots at the fiducial normalization.

\subsection{The power kernel}

The basic ingredients are the same as in
Ref.~\cite{SatoPolito:2025dist}. We bin the population in GW frequency
$f_j$ and total mass $M_k$ (the index $k$ runs over logarithmic mass
bins). A single binary of total mass $M$, mass ratio $q$, at redshift
$z$, averaged over its inclination and polarization, contributes a
characteristic strain squared
\begin{equation}
\label{eq:hs2_single}
h^2_s \propto (GM)^{10/3}\left[\frac{q}{(1+q)^2}\right]^2
\frac{(1+z)^{4/3}}{\chi^2(z)}\,
f^{4/3}\,\frac{f}{\Delta f},
\end{equation}
where $\chi(z)$ is the comoving distance and the factor $f/\Delta f$
($\Delta f = 1/T_{\rm obs}$) counts the cycles the binary spends in a
frequency bin (Ref.~\cite{SatoPolito:2025dist}, Eq.~10). A nearer binary
of given mass is louder through the explicit $(1+z)^{4/3}/\chi^2(z)$
factor.

We write $h^2_s(f_j,M_k)$ for the mean per-source value in bin $(j,k)$.
It is the $\bar N$-weighted average of Eq.~\ref{eq:hs2_single} over all
$(z,q)$ that contribute to the bin. The mean total characteristic strain
squared at frequency $f$ is then the sum over all sources,
\begin{equation}
\label{eq:hc_sum}
h_c^2(f) = \sum_k \bar{N}(f, M_k) \cdot h^2_s(f, M_k),
\end{equation}
where $\bar N(f,M_k)$ is the expected number of binaries in the bin. We
write $h_c$ for this total (the ``background'') and $h_s$ for a single
source. The velocity-dispersion-function normalization $\phi_\sigma$
multiplies every $\bar N$, so $h_c^2 \propto \phi_\sigma$ while the
per-source strain $h_s^2$ is independent of it.

\begin{figure*}
\centering
\includegraphics[width=\textwidth]{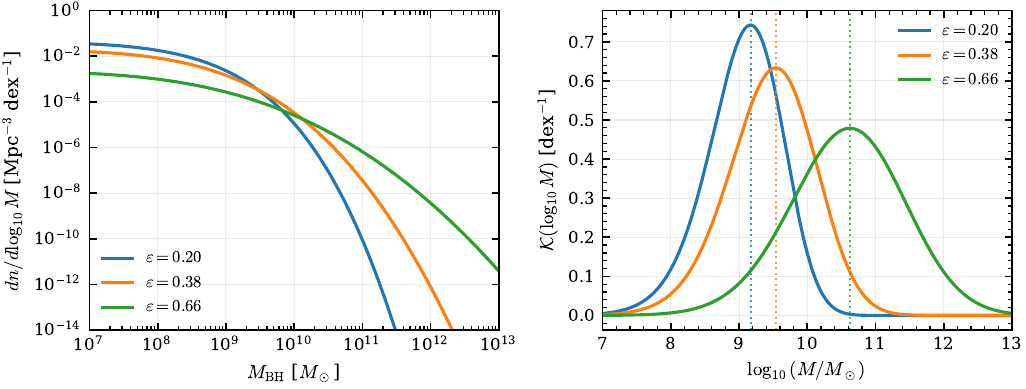}
\caption{Left: the SMBH mass function $dn/d\log_{10}M$ for three values
of the $M$--$\sigma$ scatter $\varepsilon$, each at the $\phi_\sigma$
normalization that matches the measured GWB amplitude. The low-scatter
models require an abundance well above the fiducial local value (see
text). Higher scatter populates the high-mass tail. Right: the
normalized power kernel $\mathcal{K}(\log_{10}M)$, whose integral is 1,
showing which masses contribute most to the total GW power $h_c^2$. The
dotted lines mark $M_{\rm peak}$.
\genby{scripts/fig\_mass\_function\_kernel.py}}
\label{fig:mass_kernel}
\end{figure*}

It is useful to ask which black hole masses carry the background power.
The answer is the contribution of each logarithmic mass bin to the sum
in Eq.~\ref{eq:hc_sum}, divided by the total. We call it the power
kernel,
\begin{equation}
\mathcal{K}(\log_{10} M) = \frac{\bar{N}(M) \cdot h^2_s(M)}{h_c^2}
\end{equation}
(Fig.~\ref{fig:mass_kernel}). It integrates to one by construction, so
it can be read as the distribution of the mass of the binaries that
produce the background, and its peak defines $M_{\rm peak}$, which
shifts to higher masses with increasing $\varepsilon$. The kernel does
not depend on frequency, because $\bar N$ and $h_s^2$ carry the same
$f$-scaling in every mass bin.

\subsection{Frequency grid and realizations}
\label{sec:realizations}

Throughout the paper we use the frequency grid of the NANOGrav 15-year
dataset: an observation time $T_{\rm obs} = 16.03$ yr, hence a bin width
$\Delta f = 1/T_{\rm obs} \simeq 0.062\,{\rm yr}^{-1} \simeq 2$ nHz, and
$44$ bins reaching $f_{\rm max} = 9\times10^{-8}\,{\rm Hz} \simeq
2.8\,{\rm yr}^{-1}$. Within each bin we quote the spectrum-weighted mean
frequency. The first bin has center $f = 0.085\,{\rm yr}^{-1}$. All
tables and multi-frequency figures below use three representative bins,
with centers $f = 0.085$, $0.28$, and $1.03\,{\rm yr}^{-1}$ (we label
them by these values throughout). For every model we generate $10^4$
Monte Carlo realizations.

We turn the mean population into Monte Carlo skies following the
Poisson procedure of Ref.~\cite{SatoPolito:2025dist}, extended here in
two ways: each source is placed at a position on the sphere, and each
source is assigned its own brightness. A realization is a marked
Poisson process whose intensity is the binned $\bar N(f_j, M_k)$ and
whose mark is the strain of a single binary. Each binary draws a
redshift and a mass ratio from the quadrature weights used inside the
luminosity-function integral of Ref.~\cite{SatoPolito:2023big}, and an
isotropic inclination. The inclination enters as an
orientation-dependent power factor normalized to unit mean, and the
strain of the binary is Eq.~\ref{eq:hs2_single} evaluated at the
sampled $(z,q)$ times that factor. By construction the conditional mean
per bin equals the binned value, $\mathbb{E}[h_s^2 \mid f_j, M_k] =
h_s^2(f_j, M_k)$, so the mean spectrum of Eq.~\ref{eq:hc_sum} and its
amplitude calibration are unchanged. What the sampling adds is the
within-bin source-to-source brightness scatter: a nearer, more face-on
binary is louder than a distant, edge-on one of the same mass. This
brightness scatter, together with the spread across mass bins, is what
sets the brightest-source fraction $p_1$ and the effective source count
$N_{\rm eff}$ studied below.

Each realization is stored as a ledger of bright sources plus a
remainder. The $500$ brightest binaries are drawn one by one from the
exact order-statistic tail of that process and recorded with their
strain, mass bin, and sky position. The cut at $500$ is a cut on rank,
not a threshold on the expected occupancy of a bin, so the ledger
always contains the brightest source of the realization. Everything
fainter is carried by a single compound-Poisson remainder, stored with
its count, the expectation and variance of its total power, and its
second moment $S_2 = \int w^2 \, d\Lambda$. The faint population
therefore enters through its own first two moments rather than through
its mean alone.

The sources are then scattered over a HEALPix~\cite{Gorski:2005fr} map
with $N_{\rm side} = 8$ ($N_{\rm pix} = 12 N_{\rm side}^2 = 768$ pixels of
$\approx 53.7\,{\rm deg}^2$ each, an angular resolution of
$\approx 7.3^\circ$). Each ledger source is given its own pixel, drawn
uniformly at random. The remainder is spread over the map as a positive
field that reproduces its first two pixel moments.

From each realization we form the total power $h^2_{\rm tot}$ and the
brightest-source fraction $p_1 = h^2_{s,\max}/h^2_{\rm tot}$. The
monopole and the dipole are constructed directly from the sources, not
from the map. We take $C_0$ from the total power of the realization and
$C_1/C_0 = |\sum_a p_a \hat{n}_a|^2$ from the recorded source
directions, with the unresolved remainder entering as the covariance of
an isotropic random walk. This is Eq.~\ref{eq:dipole} evaluated exactly
for the ledger, and it is the form used for the analytic laws of
Sec.~\ref{sec:dipole_source}. The map is used only for $\ell \ge 2$ and
for visualization. Its angular power spectrum, computed with
{\tt healpy}~\cite{Zonca:2019vzt} (\texttt{anafast}), is kept as a
cross-check on the dipole: the 5th, 50th, and 95th percentiles of the
two computations differ by at most
$1.9\%$.\dataref{anafast_vs_exact_pct_max_reldiff}{0.019}

\subsection{Mean versus median spectrum}
\label{sec:mean_median}

\begin{figure}
\centering
\includegraphics[width=\columnwidth]{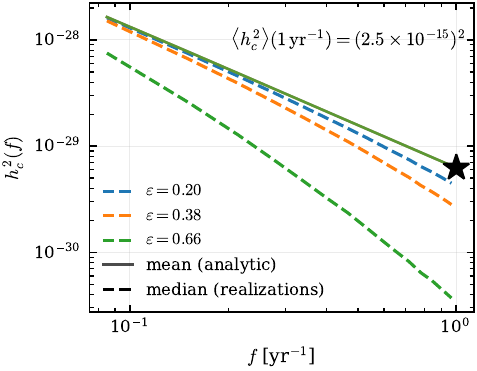}
\caption{Raw characteristic-strain spectrum $h_c^2(f)$ for three models, all
normalized to the same mean amplitude $\sqrt{\langle h_c^2\rangle} =
2.5\times10^{-15}$ at $f = 1\,{\rm yr}^{-1}$ (star). Because the mean spectrum
is $\langle h_c^2\rangle \propto f^{-4/3}$ for every model, the three mean
curves (thin solid) coincide. Their overlap is a direct check that the
per-model $\phi_\sigma$ calibration is consistent. The realization median
(thick dashed) falls below the mean by an amount that grows with $\varepsilon$
and with frequency, reaching a factor of $20$ for $\varepsilon = 0.66$ by
$f \approx 1\,{\rm yr}^{-1}$. Even at the lowest frequency the $\varepsilon =
0.66$ median is already only $0.46$ of the mean. A large mean--median split is
the direct signature of a background dominated by rare, bright sources.
\genby{scripts/fig\_strain\_spectrum\_models.py}}
\label{fig:strain_spectrum}
\end{figure}

A key diagnostic is the ratio of median to mean total strain. Because
all three models are matched to the same mean amplitude $\sqrt{\langle
h_c^2\rangle} = 2.5\times10^{-15}$ at $f = 1\,{\rm yr}^{-1}$,
Figure~\ref{fig:strain_spectrum} can show the raw spectrum $h_c^2(f)$
directly. The three mean curves (thin) coincide, a check that
the normalization is consistent, while the median curves
(thick dashed) peel away below them by a model-dependent amount. For
$\varepsilon = 0.66$ the median is only $46\%$ of the mean already at
the lowest frequency,\dataref{median_over_mean_eps066_f0085}{0.458}
because rare realizations with exceptionally
bright sources inflate the mean. For $\varepsilon = 0.20$ the ratio is
$99\%$ and the distribution is nearly
symmetric.\dataref{median_over_mean_eps020_f0085}{0.984} The gap grows with
frequency for the more anisotropic models: by $f = 1.03\,{\rm
yr}^{-1}$ the median is $0.67$, $0.41$, and $0.05$ of the mean for
$\varepsilon = 0.20$, $0.38$, and $0.66$,%
\dataref{median_over_mean_eps020_f103}{0.664}%
\dataref{median_over_mean_eps038_f103}{0.412}%
\dataref{median_over_mean_eps066_f103}{0.052}
because as sources become rare
the median falls well below the Poisson mean. This mean--median split is
a direct observable, and it is the origin of the spectral constraints of
Ref.~\cite{SatoPolito:2025dist}. We return to what it implies for the
population in Sec.~\ref{sec:discussion}.

\begin{figure*}
\centering
\includegraphics[width=\textwidth]{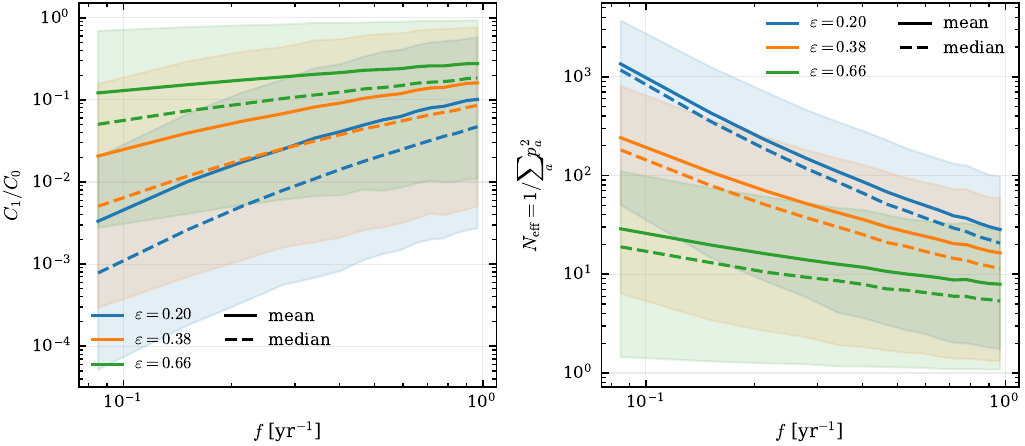}
\caption{Left: $C_1/C_0$ vs frequency for three models (mean: solid,
median: dashed, 95\% band: shading). Right: effective number of sources
$N_{\rm eff} = 1/\sum_a p_a^2$ vs frequency, computed from the full
second moment of all sources, the bright-source ledger together with the
unresolved remainder. This is the same definition used in
Table~\ref{tab:conditional} and Fig.~\ref{fig:p1_neff_conditional}
(note the logarithmic axis). Higher
$\varepsilon$ gives fewer effective sources and stronger anisotropy: at
$\varepsilon = 0.66$, the median $N_{\rm eff} \approx 5$ at $f = 1.03\,{\rm
yr}^{-1}$, rising to $\approx 19$ at the lowest frequency, whereas the
low-scatter $\varepsilon = 0.20$ model, which needs far more sources to
reach the same amplitude, has a median $N_{\rm eff}$ of order $10^3$ at the
lowest frequency. \genby{scripts/fig\_c1c0\_neff\_vs\_freq.py}}
\label{fig:c1c0_freq}
\end{figure*}

Figure~\ref{fig:c1c0_freq} shows the anisotropy statistics that follow
from the same realizations: the dipole ratio $C_1/C_0$ and the
effective source count $N_{\rm eff} = 1/\sum_a p_a^2$ as functions of
frequency. The three models are normalized to the same mean spectrum
but differ by orders of magnitude in source content. For
$\varepsilon = 0.66$ the median $N_{\rm eff}$ is $\approx 19$ at the
lowest frequency and falls to $\approx 5$ by $f = 1.03\,{\rm yr}^{-1}$,%
\dataref{neff_median_eps066_f0085}{18.9}\dataref{neff_median_eps066_f103}{5.3}
with the median dipole rising from $C_1/C_0 = 0.050$ to $0.19$.%
\dataref{c1c0_median_eps066_f0085}{0.050}\dataref{c1c0_median_eps066_f103}{0.191}
For
$\varepsilon = 0.20$ the median $N_{\rm eff}$ at the lowest frequency is
$\approx 1.2\times10^3$ and the median dipole is below
$10^{-3}$.\dataref{neff_median_eps020_f0085}{1167.0}\dataref{c1c0_median_eps020_f0085}{0.001} The
lowest frequencies, where the PTA is most sensitive, are also where
every model is at its most isotropic. The anisotropy grows toward high
frequency as the number of contributing sources drops. The effective
source count in the right panel is the reciprocal of the
per-realization shot noise
$\hat C_{\rm shot}/4\pi = \hat h_c^4/(\hat h_c^2)^2$ of
Ref.~\cite{LinLidzMa:2026b}, and where the two calculations overlap
they agree.

\section{The source content of the dipole}
\label{sec:dipole_source}

The distribution of the dipole ratio $C_1/C_0$ across realizations is
controlled by the weight fractions $\{p_a\}$, which themselves fluctuate
due to Poisson sampling and the per-source brightness scatter of
Sec.~\ref{sec:realizations}. Throughout, we write $p \equiv p_1 =
\max_a p_a$ for the brightest-source fraction. In this section we show,
in both directions, that a large dipole and a single dominant
source are equivalent statements, and then that in absolute terms
this dominant source is too faint to detect at current sensitivity.

\subsection{The distribution of $C_1/C_0$ at fixed weights}
\label{sec:forward}

The dipole is the squared norm of the weighted random walk
$\mathbf{S} = \sum_a p_a\hat n_a$ (Eq.~\ref{eq:dipole}), and simple,
accurate approximations to its distribution follow from keeping the
largest steps explicitly. When the brightest source dominates the
anisotropy budget, $p_1^2 \gg \eta \equiv \sum_{a\geq2} p_a^2$,
the remaining sources act only as an isotropic background and
\begin{equation}
\frac{C_1}{C_0} \approx p_1^2 .
\end{equation}
Keeping the brightest source and modeling the remainder as a Gaussian
faint background of variance $\eta$ makes
$\mathbf{S} = p_1\hat n + \mathbf{B}$, with $\mathbf{B} \sim
\mathcal{N}(0, \tfrac{\eta}{3}I_3)$, a three-dimensional Gaussian
centered on $p_1\hat n$. The dipole $x = C_1/C_0$ is its squared norm,
so the density of $x$ is what is left after integrating over the
direction of $\mathbf{S}$, and that angular integral is where the
$\sinh$ below comes from:
\begin{equation}
\label{eq:1src_gauss}
f(x \mid p, \eta) = \sqrt{\frac{3}{2\pi\eta}} \frac{1}{p}
\exp\!\left[-\frac{3(x + p^2)}{2\eta}\right]
\sinh\!\left(\frac{3p\sqrt{x}}{\eta}\right),
\end{equation}
which collapses to $\delta(x - p^2)$ as $\eta \to 0$. Two further
refinements keep the two brightest sources explicitly: with only two
anisotropic contributors the mutual angle makes $C_1/C_0$ exactly
uniform on $[(p_1-p_2)^2, (p_1+p_2)^2]$, and adding a Gaussian
background of variance $\eta_2 = \sum_{a\geq3}p_a^2$ to the two-source
case gives a closed form in terms of the normal CDF. Four
approximations to the distribution of the dipole are now on the table:
(i)~the brightest source alone, $\delta(x-p_1^2)$, (ii)~the uniform
two-source law, (iii)~one source plus a Gaussian background,
Eq.~\ref{eq:1src_gauss}, and (iv)~two sources plus a Gaussian
background.

\begin{figure*}
\centering
\includegraphics[width=\textwidth]{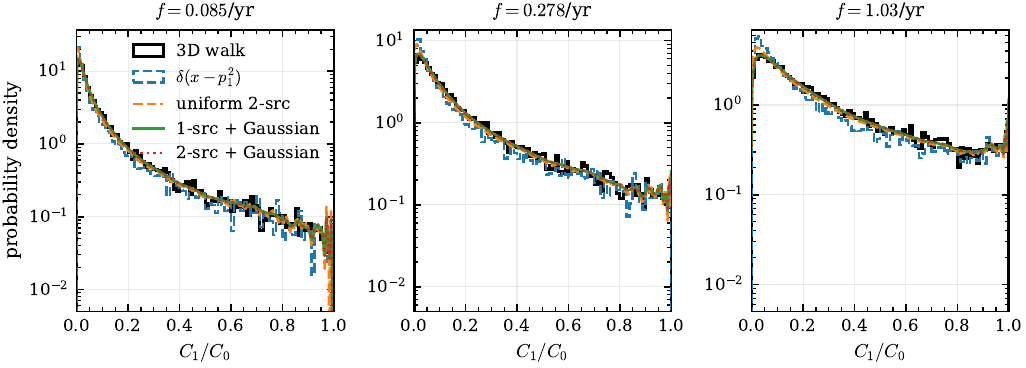}
\caption{Distribution of $C_1/C_0$ at the three representative frequencies
for the $\varepsilon = 0.66$ model ($10^4$ realizations, per-source properties
sampled). Black, labelled ``3D walk'' in the panel: the direct full-moment
walk, in which the $500$ ledger sources enter at their own directions and the
unresolved remainder enters as a Gaussian random walk of the matching second
moment. It is exact for the ledger and approximate only in that remainder.
Blue dashed: $\delta(x - p_1^2)$
(brightest source only). Orange dashed: uniform two-source. Green solid:
one-source plus Gaussian background. Red dotted: two-source plus Gaussian
background. At the lowest frequency ($f = 0.085$/yr, median
$N_{\rm eff} \approx 19$), the median is $C_1/C_0 \approx 0.05$ with a
long tail extending toward $1$. At $f = 1.03\,{\rm yr}^{-1}$ (median
$N_{\rm eff} \approx 5$), the distribution is much broader, with median
$C_1/C_0 \approx 0.19$. All four approximations capture the shape well, and the
one-source plus Gaussian formula (Eq.~\ref{eq:1src_gauss}) is the closest of
them. \genby{scripts/mc/generate\_sampled\_c1c0\_distribution.py}}
\label{fig:c1c0_pdf}
\end{figure*}

All four are evaluated per realization and superimposed on the direct
full-moment walk in Fig.~\ref{fig:c1c0_pdf}. What matters there is not
which approximation is most accurate but that the crudest one already
works: keeping only the brightest source, $\delta(x-p_1^2)$, reproduces
the shape of the full distribution at every frequency. The dipole of
one of these skies is set by a single weight, and the rest of the
population only fills in the detail. Adding that detail as a Gaussian
background, Eq.~\ref{eq:1src_gauss}, tracks the full distribution to
within a few percent.

\begin{figure*}
\centering
\includegraphics[width=\textwidth]{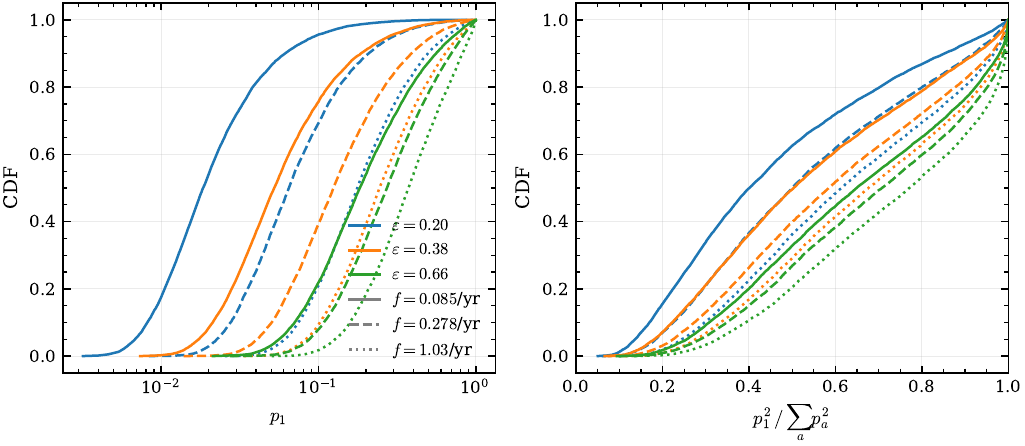}
\caption{Left: CDF of the brightest-source fraction $p_1$ for three
$\varepsilon$ models at three frequencies (solid: $f = 0.085$ yr$^{-1}$,
dashed: $f = 0.278$ yr$^{-1}$, dotted: $f = 1.03$ yr$^{-1}$).
Right: fraction of the conditional mean $\langle C_1/C_0 \rangle
= \sum_a p_a^2$ contributed by $p_1^2$ alone, using the full second moment
of all sources. The concentration is strongest at $\varepsilon = 0.66$
(green), where the median runs from $0.650$ at the lowest frequency to
$0.771$ at $f = 1.03\,{\rm yr}^{-1}$,%
\dataref{p1sq_fraction_median_eps066_f0085}{0.650}%
\dataref{p1sq_fraction_median_eps066_f103}{0.771}
and weakest at $\varepsilon = 0.20$ (blue), where it runs from $0.401$ to
$0.614$ over the same range.%
\dataref{p1sq_fraction_median_eps020_f0085}{0.401}%
\dataref{p1sq_fraction_median_eps020_f103}{0.614}
\genby{scripts/fig\_p1\_distributions.py}}
\label{fig:p1_distributions}
\end{figure*}

Figure~\ref{fig:p1_distributions} shows why. Its left panel is the
distribution of the brightest-source fraction $p_1$. Its right panel is
the fraction $p_1^2/\sum_a p_a^2$ of the direction-averaged dipole
power carried by that one source. For $\varepsilon = 0.66$ the median
realization carries $65\%$ of that power in the brightest source at the
lowest frequency, rising to $77\%$ at $f = 1.03\,{\rm yr}^{-1}$ where
fewer sources contribute.%
\dataref{p1sq_fraction_median_eps066_f0085}{0.650}%
\dataref{p1sq_fraction_median_eps066_f103}{0.771}

This establishes one direction of the argument. Producing a large dipole
requires a dominant source, because many comparable weights added in
random directions average down to a small $C_1/C_0$. The two figures are
what support that statement: the distribution of the dipole is
reproduced by the brightest source alone, and the brightest source
carries most of the direction-averaged dipole power in the median
realization. The converse, that a large measured dipole implies a
dominant source, is shown next.

\subsection{The source content of a given dipole}
\label{sec:inverse_c1c0}

The complementary and more directly useful question is the inverse:
given a measured dipole $C_1/C_0 = x$, what does it imply about
the source population? Two summaries answer this: the brightest-source
fraction $p_1$ and the effective number of sources $N_{\rm eff} =
1/\sum_a p_a^2$ (Eq.~\ref{eq:neff}).

By Bayes' theorem the distribution of $p_1$ given that the dipole exceeds
a measured value $v$ is
\begin{equation}
\label{eq:p1_given_x}
P(p_1 \mid C_1/C_0 > v) \propto
   \pi(p_1)\int d\eta\; S(v \mid p_1, \eta)\,P(\eta \mid p_1),
\end{equation}
where $S(v\mid p_1,\eta) \equiv P(C_1/C_0 > v\mid p_1,\eta)$ is the
exceedance probability under Eq.~\ref{eq:1src_gauss}, $\pi(p_1)$ is the
population prior on the brightest-source fraction, and
$P(\eta\mid p_1)$ its conditional spread.
We have no closed form for the priors: they are properties of the
Poisson-sampled population, so we take them directly from the
simulation and evaluate Eq.~\ref{eq:p1_given_x} by importance
reweighting. Each realization is reweighted by its exceedance
probability $S(v\mid p_1,\eta)$, and the reweighted histogram of $p_1$
(or of $N_{\rm eff}$) is the prediction. This reproduces the directly
conditioned Monte Carlo almost exactly
(Fig.~\ref{fig:p1_neff_conditional}).

\begin{figure*}
\centering
\includegraphics[width=\textwidth]{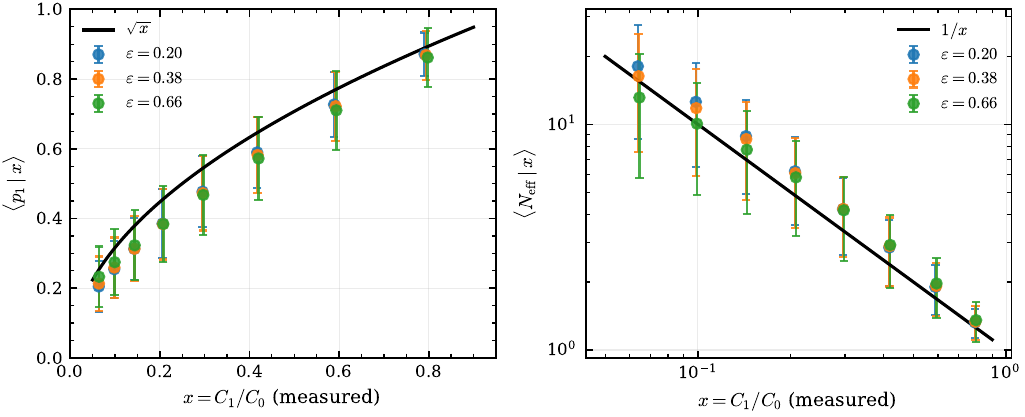}
\caption{The source content of a dipole is universal across the cases
tested, the three calibrated models and the full frequency grid of
Sec.~\ref{sec:realizations}. Realizations of all
three models and all frequencies are pooled and binned by their
measured $C_1/C_0 = x$. Points show the mean (with $1\sigma$ scatter) of the
brightest-source fraction $p_1$ (left) and the effective number of sources
$N_{\rm eff}=1/\sum_a p_a^2$ (right) in each bin. The simple closed forms
$\langle p_1\mid x\rangle \simeq \sqrt{x}$ (Eq.~\ref{eq:p1_given_x_simple})
and $\langle N_{\rm eff}\mid x\rangle \simeq 1/x$ (Eq.~\ref{eq:neff_given_x},
solid black) track the Monte Carlo across the whole range, and the three
models collapse onto the same relation regardless of frequency: what a dipole
implies about the source population depends on the dipole, not on the model
or the frequency. \genby{scripts/fig\_source\_content\_estimators.py}}
\label{fig:source_content_simple}
\end{figure*}

\begin{figure*}
\centering
\includegraphics[width=\textwidth]{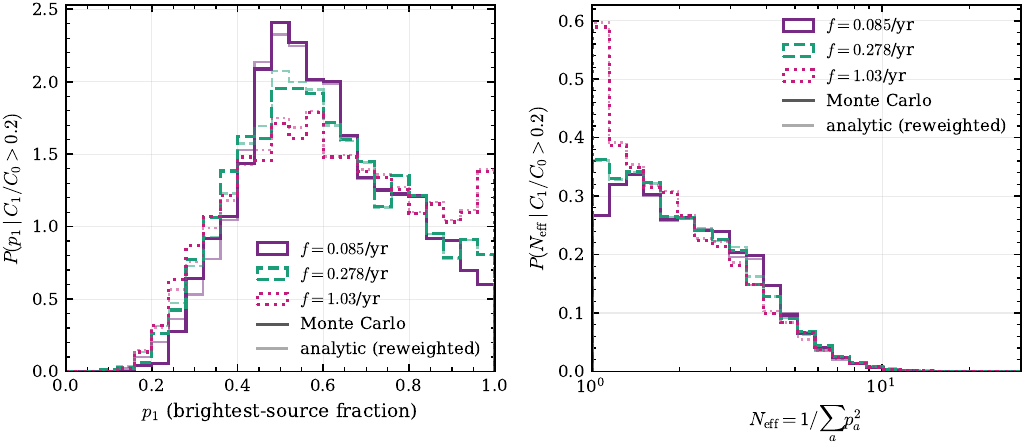}
\caption{Source content of a large dipole for the $\varepsilon = 0.66$ model,
at the three representative frequencies, conditioned on
$C_1/C_0 > 0.2$ (the published NANOGrav 95\% value). Left: distribution of the
brightest-source fraction $p_1$; right: distribution of the effective number of
sources $N_{\rm eff} = 1/\sum_a p_a^2$. Color and line style both encode
frequency, and shade encodes method (dark: Monte Carlo; light: the analytic
prediction of Eq.~\ref{eq:p1_given_x}, obtained by importance reweighting each
Monte Carlo realization by its exceedance probability $S(0.2\mid p_1, \eta)$
under Eq.~\ref{eq:1src_gauss}). Because this figure shows a single model, its
colors encode frequency rather than $\varepsilon$, unlike every other figure
here. A dipole above the published value requires
the brightest source to carry $\sim 60\%$ of the power, with only $\sim 3$
effective sources, nearly independent of model and frequency
(cf.\ Fig.~\ref{fig:source_content_simple}). \genby{scripts/fig\_p1\_neff\_conditional.py}}
\label{fig:p1_neff_conditional}
\end{figure*}

Two back-of-envelope relations summarize how a measured dipole $x$ pins
down the source content. When the brightest source dominates,
Eq.~\ref{eq:1src_gauss} collapses to $\delta(x-p_1^2)$, so
\begin{equation}
\label{eq:p1_given_x_simple}
\langle p_1 \mid x\rangle \simeq \sqrt{x},
\end{equation}
lowered slightly by the sub-dominant sources. And since the direction
average of the dipole is $\langle C_1/C_0 \mid \{p_a\}\rangle = \sum_a
p_a^2 = 1/N_{\rm eff}$ (Eq.~\ref{eq:mean_cl}), a measured $x$ lets us
estimate $1/N_{\rm eff}$, up to realization scatter,
\begin{equation}
\label{eq:neff_given_x}
\langle N_{\rm eff} \mid x\rangle \simeq 1/x .
\end{equation}
Figure~\ref{fig:source_content_simple} tests both against the Monte
Carlo. Pooling every realization at every frequency and binning by the
measured $x$, the conditional means follow $\sqrt{x}$ and $1/x$ closely,
and the three models and all frequencies fall on the same curves. The
source content implied by a dipole is set by the dipole itself, not by
the underlying model or by the frequency at which it is measured. Those
only control how often a given dipole occurs.
Figure~\ref{fig:p1_neff_conditional} shows the full conditional
distributions behind these means, for the $\varepsilon = 0.66$ model at
the three representative frequencies, conditioned on
$C_1/C_0 > 0.2$, the value quoted by the NANOGrav anisotropy
analysis, whose meaning we examine in Sec.~\ref{sec:nanograv}.

Table~\ref{tab:conditional} quantifies this for all three models and the
three representative frequencies of Sec.~\ref{sec:realizations}. The
conditional $p_1$ and $N_{\rm eff}$ are set essentially by the
threshold alone, not by the model or frequency, while the exceedance
probability varies by orders of magnitude. The faint-background variance
$\eta$ is small everywhere but not constant: across the
$C_1/C_0 > 0.2$ cells its mean runs from $0.004$ to $0.028$.%
\dataref{eta_mean_given_02_min}{0.004}\dataref{eta_mean_given_02_max}{0.028}
The reason is built into
Eq.~\ref{eq:1src_gauss}. Conditioning on a large dipole selects
realizations in which $C_1/C_0 \approx p_1^2 + \eta$ is large. Since
the faint background carries $\eta \ll 1$, the exceedance probability
acts as a near-step selection on $p_1$ at $\sqrt{v}$, almost
independently of the population prior. The prior sets how many
realizations clear the threshold. In $p_1$ and $N_{\rm eff}$, the
realizations that do are all alike.

\begin{table*}
\centering
\caption{What a given dipole implies. For three models, the three
representative frequencies of Sec.~\ref{sec:realizations}, and three
thresholds, the probability that $C_1/C_0$ exceeds the threshold and, for the
realizations that do, the mean brightest-source fraction $p_1$, effective
source count $N_{\rm eff} = 1/\sum_a p_a^2$, and faint-background variance
$\eta = \sum_{a\geq2} p_a^2$ ($\pm 1\sigma$ on $p_1$, $N_{\rm eff}$; from $10^4$
realizations). Entries marked ``---'' have fewer than $20$ exceeding
realizations. Whenever the dipole is large the brightest source dominates
($\eta \ll p_1^2$) and only a handful of sources contribute, regardless of
model or frequency. What the model and frequency control is how often
the dipole is large. \genby{scripts/table\_conditional\_stats.py}}
\label{tab:conditional}
\scriptsize
\setlength{\tabcolsep}{3.5pt}
\begin{tabular}{lcccccccccccc}
\hline\hline
 & \multicolumn{4}{c}{$C_1/C_0 > 0.1$} & \multicolumn{4}{c}{$C_1/C_0 > 0.2$} & \multicolumn{4}{c}{$C_1/C_0 > 0.5$} \\
 & $P$ & $\langle p_1\rangle$ & $\langle N_{\rm eff}\rangle$ & $\langle\eta\rangle$ & $P$ & $\langle p_1\rangle$ & $\langle N_{\rm eff}\rangle$ & $\langle\eta\rangle$ & $P$ & $\langle p_1\rangle$ & $\langle N_{\rm eff}\rangle$ & $\langle\eta\rangle$ \\
\hline
\multicolumn{13}{l}{\textit{$\varepsilon = 0.20$}} \\
\quad $f = 0.085$ & 0.004 & $0.41\pm0.12$ & $6.7\pm2.8$ & $0.003$ & 0.001 & --- & --- & --- & 0.000 & --- & --- & --- \\
\quad $f = 0.28$ & 0.050 & $0.44\pm0.16$ & $6.4\pm3.8$ & $0.006$ & 0.022 & $0.58\pm0.14$ & $3.3\pm1.3$ & $0.005$ & 0.004 & $0.81\pm0.07$ & $1.6\pm0.3$ & $0.001$ \\
\quad $f = 1.03$ & 0.301 & $0.45\pm0.19$ & $6.5\pm4.8$ & $0.019$ & 0.149 & $0.58\pm0.17$ & $3.4\pm1.9$ & $0.015$ & 0.040 & $0.80\pm0.10$ & $1.6\pm0.4$ & $0.006$ \\
\multicolumn{13}{l}{\textit{$\varepsilon = 0.38$}} \\
\quad $f = 0.085$ & 0.043 & $0.45\pm0.15$ & $6.1\pm3.3$ & $0.005$ & 0.018 & $0.59\pm0.13$ & $3.2\pm1.3$ & $0.004$ & 0.004 & $0.79\pm0.08$ & $1.6\pm0.3$ & $0.001$ \\
\quad $f = 0.28$ & 0.175 & $0.45\pm0.18$ & $6.4\pm4.4$ & $0.012$ & 0.085 & $0.59\pm0.16$ & $3.4\pm1.7$ & $0.009$ & 0.020 & $0.80\pm0.10$ & $1.6\pm0.4$ & $0.003$ \\
\quad $f = 1.03$ & 0.467 & $0.47\pm0.21$ & $5.9\pm4.5$ & $0.024$ & 0.271 & $0.59\pm0.19$ & $3.4\pm2.1$ & $0.020$ & 0.081 & $0.81\pm0.12$ & $1.6\pm0.5$ & $0.009$ \\
\multicolumn{13}{l}{\textit{$\varepsilon = 0.66$}} \\
\quad $f = 0.085$ & 0.324 & $0.48\pm0.20$ & $5.9\pm4.5$ & $0.018$ & 0.180 & $0.61\pm0.18$ & $3.2\pm1.8$ & $0.015$ & 0.056 & $0.81\pm0.10$ & $1.6\pm0.4$ & $0.006$ \\
\quad $f = 0.28$ & 0.514 & $0.49\pm0.22$ & $5.5\pm4.2$ & $0.025$ & 0.312 & $0.60\pm0.19$ & $3.3\pm2.1$ & $0.021$ & 0.106 & $0.81\pm0.12$ & $1.6\pm0.5$ & $0.009$ \\
\quad $f = 1.03$ & 0.692 & $0.53\pm0.23$ & $4.6\pm3.6$ & $0.033$ & 0.484 & $0.62\pm0.21$ & $3.1\pm2.0$ & $0.028$ & 0.196 & $0.81\pm0.13$ & $1.6\pm0.5$ & $0.013$ \\
\hline\hline
\end{tabular}
\end{table*}

At the lowest frequency, where the PTA is actually sensitive to anisotropy
(Sec.~\ref{sec:source_detection}), reaching $C_1/C_0 = 0.2$ is
appreciable only for the most anisotropic model: $P(C_1/C_0 > 0.2) = 0.18$
for $\varepsilon = 0.66$, versus $0.018$ for $\varepsilon = 0.38$ and
$0.001$ for $\varepsilon = 0.20$.%
\dataref{p_exceed_02_eps066_f0085}{0.180}%
\dataref{p_exceed_02_eps038_f0085}{0.018}%
\dataref{p_exceed_02_eps020_f0085}{0.001} And even in that single favorable case,
the highest $\varepsilon$ at the lowest frequency, exceeding $0.2$
requires the brightest source to carry $\langle p_1\rangle = 0.61 \pm
0.18$ of the power, with only $\langle N_{\rm eff}\rangle = 3.2 \pm 1.8$
effective sources.%
\dataref{p1_mean_given_02_eps066_f0085}{0.61}%
\dataref{neff_mean_given_02_eps066_f0085}{3.2} A measured dipole at that level would not be a diffuse
anisotropy of the background. It would be the signature of one or two
individually loud binaries. For the astrophysically expected signal,
measuring $C_\ell$ is therefore an indirect way of measuring the single
brightest source. That raises the question of whether the source can be
seen at all.

\subsection{The brightest source in absolute terms}
\label{sec:brightest_absolute}

The fraction $p_1$ is a relative measure: its denominator is the total
power $h^2_{\rm tot}$ of the same realization, which itself fluctuates.
We therefore define a complementary quantity with a fixed reference, the
ratio of the brightest source's power to the model's mean background,
\begin{equation}
\label{eq:R_ratio}
R \equiv \frac{h^2_{s,\max}}{\langle h_c^2\rangle(f)},
\end{equation}
whose denominator is the same for every realization and equals the observed
amplitude by construction (Sec.~\ref{sec:astro}). Unlike $p_1$, $R$ can be
compared directly to what a continuous-wave search constrains: the brightest
source's strain amplitude $h_0 = \sqrt{h^2_{s,\max}/(f\,T_{\rm obs})}$ (using
$h_s^2 = h_0^2\,f/\Delta f$, with $\Delta f = 1/T_{\rm obs}$ and $T_{\rm obs}
= 16.03$ yr) set against the mean background amplitude $A(f) =
\sqrt{\langle h_c^2\rangle(f)}$.
Table~\ref{tab:source_ratio} lists the median and $95$th percentile of $R$ and
of $h_0$. The median brightest source carries only a small fraction of the
background ($R \lesssim 0.12$), but the heavy tail means that in the loudest
realizations it rivals the entire background ($R \to 1$). Translated to a
strain amplitude, the $95$th-percentile $h_0$ at the lowest frequency is
$3.5\times10^{-15}$, $6.6\times10^{-15}$ and $1.2\times10^{-14}$ for
$\varepsilon = 0.20$, $0.38$ and $0.66$.%
\dataref{h0_95pct_eps020_f0085}{3.5e-15}\dataref{h0_95pct_eps038_f0085}{6.6e-15}%
\dataref{h0_95pct_eps066_f0085}{1.17e-14}

\begin{figure*}
\centering
\includegraphics[width=\textwidth]{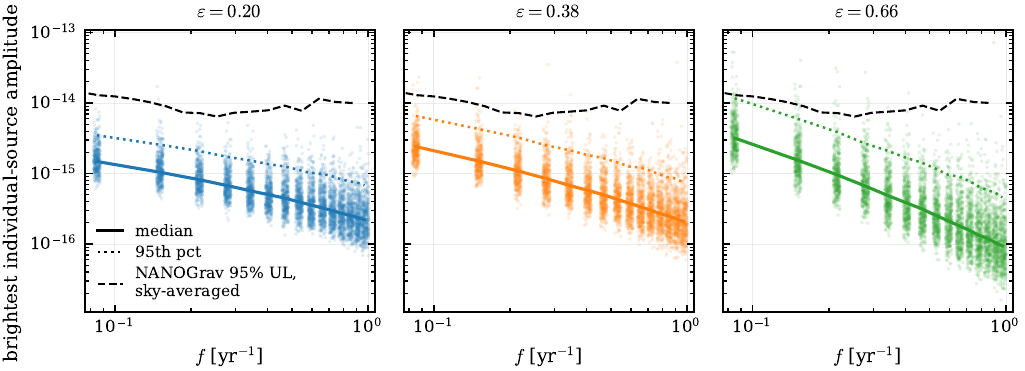}
\caption{Distribution of the brightest individual source in each Monte Carlo
realization, expressed as a strain amplitude $h_0 =
\sqrt{h_s^2/(f\,T_{\rm obs})}$, versus GW frequency, for the three models
(dots: a subsample of realizations; solid: median; dotted: 95th percentile).
The dashed black curve is the NANOGrav 15-year $95\%$ upper limit on an
individual continuous-wave source~\cite{NANOGrav:2023individual}, averaged
over the $192$ sky pixels of the published limit grid. It is a sky-averaged
bound, not a position-independent one: at the lowest frequency plotted the
per-pixel limit runs from $5.2\times10^{-15}$ to
$2.7\times10^{-14}$%
\dataref{cw_limit_perpixel_min_f0085}{5.2e-15}%
\dataref{cw_limit_perpixel_max_f0085}{2.7e-14}
about the sky mean. The curve is drawn only where that grid has entries and
stops at $f = 0.87\,{\rm yr}^{-1}$.\dataref{cw_limit_fmax_per_year}{0.868}
Every model lies below it at every frequency, with the $\varepsilon = 0.66$
95th percentile coming closest at the low-frequency end.
\genby{scripts/fig\_pixel\_amplitude.py}}
\label{fig:pixel_amplitude}
\end{figure*}

Figure~\ref{fig:pixel_amplitude} sets the brightest individual source of
every realization against the NANOGrav 15-year continuous-wave upper
limit on $h_0$, averaged over the
sky~\cite{NANOGrav:2023individual}. That limit is not flat. It is
deepest in the middle of the band and weaker at both ends, and at the
lowest frequency we consider, where the loudest sources sit, it is
$1.3\times10^{-14}$.\dataref{cw_limit_skyavg_f0085}{1.30e-14} No model
reaches it at any frequency the published grid covers. The closest
approach is the $95$th percentile of the $\varepsilon = 0.66$ model at
the lowest frequency, which is $0.90$ of the
limit.\dataref{h0_over_cwlimit_max95_eps066}{0.90} The same maximum is
$0.50$ for $\varepsilon = 0.38$ and $0.29$ for $\varepsilon = 0.20$,%
\dataref{h0_over_cwlimit_max95_eps038}{0.50}%
\dataref{h0_over_cwlimit_max95_eps020}{0.29}
and the median brightest source stays below $0.24$ of the limit in every
model.\dataref{h0_over_cwlimit_maxmedian_eps066}{0.24} Individual
realizations scatter about the median (dots) by roughly an order of
magnitude, so the loudest realizations of the lower-scatter models climb
toward the limit even though their medians sit far below it.

Two things follow. The anisotropy of the astrophysically expected
background is one bright binary on an otherwise nearly isotropic
background. In every model normalized to the observed amplitude, that
binary is individually too faint for current continuous-wave
sensitivity. Whatever statistic one uses to chase the anisotropy, the
underlying object being chased is a source that current data cannot
detect.

\begin{table}
\centering
\caption{The brightest source in absolute terms. For each model and frequency,
the median and $95$th percentile of the brightest-to-background power ratio
$R = h^2_{s,\max}/\langle h_c^2\rangle$ (Eq.~\ref{eq:R_ratio}) and of the
brightest source's strain amplitude $h_0$, from $10^4$ realizations. The mean
background amplitude is $A = \sqrt{\langle h_c^2\rangle} = 1.3\times10^{-14}$,
$5.9\times10^{-15}$, $2.45\times10^{-15}$ at $f = 0.085, 0.28, 1.03\,{\rm
yr}^{-1}$. For comparison, the NANOGrav 15-year $95\%$ upper limit on an
individual continuous-wave source, averaged over the $192$ sky positions of
the published limit grid~\cite{NANOGrav:2023individual}, is
$1.3\times10^{-14}$ at $f = 0.085\,{\rm
yr}^{-1}$.\dataref{cw_limit_skyavg_f0085}{1.30e-14} That limit is
frequency-dependent, and Fig.~\ref{fig:pixel_amplitude} shows the whole
curve. \genby{scripts/table\_conditional\_stats.py}}
\label{tab:source_ratio}
\footnotesize
\begin{tabular}{lcccc}
\hline\hline
 & median $R$ & 95\% $R$ & median $h_0$ & 95\% $h_0$ \\
\hline
\multicolumn{5}{l}{\textit{$\varepsilon = 0.20$}} \\
\quad $f = 0.085$ & 0.018 & 0.101 & $1.5\times10^{-15}$ & $3.5\times10^{-15}$ \\
\quad $f = 0.28$ & 0.059 & 0.384 & $6.8\times10^{-16}$ & $1.7\times10^{-15}$ \\
\quad $f = 1.03$ & 0.117 & 1.216 & $2.1\times10^{-16}$ & $6.7\times10^{-16}$ \\
\multicolumn{5}{l}{\textit{$\varepsilon = 0.38$}} \\
\quad $f = 0.085$ & 0.048 & 0.358 & $2.4\times10^{-15}$ & $6.6\times10^{-15}$ \\
\quad $f = 0.28$ & 0.093 & 0.818 & $8.5\times10^{-16}$ & $2.5\times10^{-15}$ \\
\quad $f = 1.03$ & 0.100 & 1.456 & $1.9\times10^{-16}$ & $7.3\times10^{-16}$ \\
\multicolumn{5}{l}{\textit{$\varepsilon = 0.66$}} \\
\quad $f = 0.085$ & 0.084 & 1.137 & $3.2\times10^{-15}$ & $1.2\times10^{-14}$ \\
\quad $f = 0.28$ & 0.057 & 0.923 & $6.7\times10^{-16}$ & $2.7\times10^{-15}$ \\
\quad $f = 1.03$ & 0.019 & 0.527 & $8.4\times10^{-17}$ & $4.4\times10^{-16}$ \\
\hline\hline
\end{tabular}
\end{table}

\section{Searching for the source}
\label{sec:source_detection}

Section~\ref{sec:dipole_source} established that the anisotropy expected
from the astrophysical background is carried by the single brightest
source, and that this source is faint in absolute terms. The operational
question is then: if a source carries a fraction $p_1$ of a background
that has itself been measured with some signal-to-noise ratio, what is
the best way to detect it? And how much is lost by compressing the
anisotropy to an angular power spectrum $C_\ell$?

A source in a known direction stamps a known pattern on the
pulsar-pair data. The best way to look for a known pattern in noisy
data is to correlate the data against that pattern, weighting by the
inverse noise. That operation is a matched filter, and it is what a
continuous-wave search performs. We compare it against the $C_\ell$
statistic because $C_\ell$ is what the published anisotropy analyses
measure, and because both are built from the same pair data.

The section proceeds in four steps. Section~\ref{sec:sd_setup} fixes
what the data are and what the background significance $\rho_0$ means.
Section~\ref{sec:sd_coherent} derives the significance of the matched
filter. Section~\ref{sec:sd_cl} compares that with the $C_\ell$
statistic at the same angular resolution. Section~\ref{sec:sd_bottom}
draws the consequences for the current data and for a future array. The
answer explains why no source, and hence no anisotropy, is close to
being detected in current data.
\provnote{The analytic numbers in this section are reproduced by
\texttt{scripts/verify\_source\_vs\_cl.py} and unit-tested in
\texttt{tests/test\_source\_detection.py}; the Monte Carlo scan
penalties are those printed by
\texttt{scripts/fig\_cl\_vs\_scan\_penalty.py}.}

\subsection{The data, and what \texorpdfstring{$\rho_0$}{rho0} measures}
\label{sec:sd_setup}

Work in a single frequency bin. The raw observables are the timing residuals of
each pulsar. The quantities that carry angular information are the
cross-correlations between pulsars, one number for each pair
$(a,b)$. An isotropic gravitational-wave background of amplitude $A$ contributes
to every pair an amount fixed by the Hellings--Downs curve,
\begin{equation}
\label{eq:HD}
\Gamma_0(\zeta_{ab}) = \tfrac13 - \tfrac{u}{6} + u\ln u,
\qquad u = \tfrac{1-\cos\zeta_{ab}}{2},
\end{equation}
which depends only on the angle $\zeta_{ab}$ between the two pulsars. We
normalize $\Gamma_0$ so that it is exactly the average of the
single-source pair response of Eq.~\ref{eq:gamma_src} below over source
directions, which is the convenient choice for comparing the two
templates. The conventional Hellings--Downs curve is $\tfrac32\Gamma_0$,
and all significance ratios below are independent of this choice.
Stacking the pairs into a vector, the data covariance is
$\bar C_0 = N + A\,\Gamma_{\rm iso}$, where $N$ is the per-pulsar noise
covariance and $\Gamma_{\rm iso}$ is the Hellings--Downs pattern laid out
across all pairs.

The background amplitude $A$ is itself measured with finite precision.
Its signal-to-noise ratio sets the scale of every detection statement
below, and we write it as
\begin{equation}
\label{eq:sigma_bg}
\rho_0 \equiv \Sigma_{\rm bg}, \qquad \Sigma_{\rm bg}^2 = A^2 F_A ,
\end{equation}
where $F_A = \tfrac12\,\mathrm{Tr}[\bar C_0^{-1}\Gamma_{\rm iso}\bar C_0^{-1}
\Gamma_{\rm iso}]$ is the Fisher information on $A$. For the idealized
array we use for estimates below (equal-noise pulsars, roughly
uniform on the sky, in the noise-dominated regime) the cross-pair
contribution to $F_A$ is proportional to the pair-averaged square of the
Hellings--Downs curve, $\langle\Gamma_0^2\rangle =
\tfrac12\int_{-1}^{1}\Gamma_0(\mu)^2\,d\mu = \tfrac{1}{108}$ in the
normalization of Eq.~\ref{eq:HD}.
The size of $\rho_0$ must be anchored, because every detection statement
below scales with it. Two points matter. (i)~Anisotropy lives
entirely in the cross-correlations: the auto-power of an
individual pulsar carries no directional information, so the relevant
$\rho_0$ is the significance of the Hellings--Downs
cross-correlation, not the much larger significance of the common
red-spectrum process. (ii)~For the NANOGrav 15-year data that
cross-correlation has a total optimal-statistic signal-to-noise ratio of
$\approx 5$~\cite{NANOGrav:2023gor}, added in quadrature over all pulsar
pairs and all frequencies. This is the amplitude of the correlated signal
in units of its own uncertainty, $5\pm1$, and is a different quantity from
the $3$--$4\sigma$ significance usually quoted for the detection, which
measures Hellings--Downs correlation against a spatially uncorrelated
common-spectrum alternative rather than against zero. Roughly five
low-frequency bins carry most of
that signal, so the per-bin cross-correlation significance is
$5/\sqrt{5} \approx 2$. Everything below is written in terms of the
per-bin $\rho_0$, and we say so explicitly where the band-integrated
value is meant instead.

\subsection{The coherent search and its signal-to-noise}
\label{sec:sd_coherent}

A source in a known direction $\hat s$ produces a definite, predictable
pattern across the pulsar pairs. The response of pulsar $a$ to a wave from
$\hat s$ is captured by its two antenna-pattern functions $F_a^+(\hat s)$ and
$F_a^\times(\hat s)$, and the cross-correlation a single source induces
between pulsars $a$ and $b$ is the product of their responses, summed over
polarizations,
\begin{equation}
\label{eq:gamma_src}
\gamma_{ab}(\hat s) = F_a^+ F_b^+ + F_a^\times F_b^\times .
\end{equation}
This is, for a point source, the analogue of the Hellings--Downs curve
for the isotropic background: it is the shape the source stamps on the
pair data. Because that shape is known once $\hat s$ is fixed, the
optimal way to detect the source is a matched filter: correlate
the measured pair data against the template $\gamma_{ab}(\hat s)$,
weighted by inverse noise, exactly what a continuous-wave search
does. The isotropic background is a nuisance for this purpose. Since it
is the monopole of the power sky, fitting it out amounts to projecting
the monopole out of the template.

The source significance then takes a simple form,
\begin{equation}
\label{eq:rho_ps}
\rho_{\rm ps} = \alpha_{\rm ps}\,p_1\,\rho_0,
\qquad \alpha_{\rm ps} = \sqrt{5} \simeq 2.24 ,
\end{equation}
in the equal-noise, uniformly-distributed array. That array is the ideal
benchmark for every numerical estimate in this section, and
$\alpha_{\rm ps}^2 = 5$ holds only there. Here $\rho_{\rm ps}$ is the
noncentrality coordinate of the source pair-pattern test, and $\rho_0$
is the per-bin background significance of
Sec.~\ref{sec:sd_setup}, not the band-integrated value quoted in the
abstract. A continuous-wave pipeline that works in the timing-residual
mean defines the same test in a different numerical signal-to-noise
coordinate. The factor $p_1\rho_0$ is
intuitive: a source carrying a fraction $p_1$ of a background whose amplitude is
known to a significance $\rho_0$ has its own amplitude known to $\sim p_1\rho_0$.

The coefficient $\alpha_{\rm ps}$ is purely geometric: the size of the
source's pair-pattern $\gamma_{ab}$ relative to the background's pattern
$\Gamma_0$, once the monopole the two share has been removed. Placing the
source on the $z$-axis, the antenna patterns are
$F^+=\tfrac12(1+\mu)\cos2\phi$ and $F^\times=\tfrac12(1+\mu)\sin2\phi$.
The mean square of the source pattern over random pulsar pairs is
$\langle\gamma_s^2\rangle = \tfrac{1}{18}$, while
$\langle\Gamma_0^2\rangle = \tfrac{1}{108}$, a ratio of $6$. The
source-direction average of $\gamma_{ab}(\hat s)$ is exactly
$\Gamma_0(\zeta_{ab})$ in the normalization of Eq.~\ref{eq:HD}, so the
source pattern contains one unit of monopole, which is absorbed when the
background amplitude is fitted out, leaving
$\alpha_{\rm ps}^2 = 6 - 1 = 5$. Fitting the background away thus costs
only $\sqrt{5/6}\simeq0.91$ in SNR.

The scaling is now concrete. If the background were eventually measured at
$\rho_0=20$ in a bin, a source carrying $p_1=0.1$ of it would be found with
significance $\rho_{\rm ps}\simeq\sqrt5\times0.1\times20\simeq4.5$, a
clear detection. At today's per-bin $\rho_0 \approx 2$, the same source
gives $\rho_{\rm ps}\lesssim1$. Even the dominant source that a dipole at
$C_1/C_0 = 0.2$ would demand, $p_1 \sim 0.6$
(Sec.~\ref{sec:dipole_source}), would only reach $\rho_{\rm ps} \sim
1.3\,\rho_0$, a marginal significance at current sensitivity.

\begin{figure}
\centering
\includegraphics[width=\columnwidth]{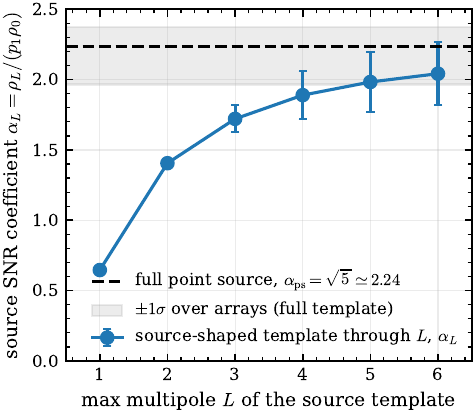}
\caption{Source detection SNR retained by an anisotropy search of angular
resolution $L$, in the idealized equal-noise array. The matched filter for the
source's full cross-correlation pattern reaches
$\rho_{\rm ps}=\alpha_{\rm ps}\,p_1\rho_0$ with
$\alpha_{\rm ps}=\sqrt5\simeq2.24$ (dashed). Truncating the template at multipole
$L$ keeps only $\alpha_L<\alpha_{\rm ps}$ (points, mean $\pm$ scatter over random
arrays of $N_p=120$ pulsars). The gray band is the corresponding $\pm1\sigma$
scatter of the full-template coefficient over the same arrays. A dipole-only
search retains about $29\%$ of the
amplitude. \genby{scripts/fig\_source\_snr\_vs\_resolution.py}}
\label{fig:source_snr_res}
\end{figure}

Equation~\ref{eq:rho_ps} uses the source's entire angular pattern.
Restricting the template to multipoles $\ell\le L$ gives a smaller coefficient,
$\rho_L = \alpha_L\,p_1\,\rho_0$ with $\alpha_L<\alpha_{\rm ps}$.
Figure~\ref{fig:source_snr_res} shows $\alpha_L$ rising from
$\alpha_1\simeq0.65$, a dipole-only search in arrays of $\gtrsim100$
pulsars, through $\alpha_{\ell\le2}\simeq1.4$, toward the full
point-source value $\sqrt5$. A dipole-only search therefore keeps only
about $29\%$ of the source amplitude, because a point source has
substantial quadrupole and higher-order structure in its pair pattern.%
\dataref{alpha_1_Np120}{0.646}\dataref{alpha_le2_Np120}{1.406}%
\dataref{dipole_amplitude_fraction}{0.289}

\subsection{Comparison with the angular power spectrum}
\label{sec:sd_cl}

The angular power spectrum keeps only the rotationally invariant block powers
$C_L^P = (2L+1)^{-1}\sum_m |a_{Lm}|^2$ of the source-power sky and discards
the $m$-pattern, the relative phases of the $a_{Lm}$, which is precisely
the information that locates the source on the sky. For a single source the
spectrum is flat in $\ell$, $C_L^P/C_0^P = p_1^2$ for $L>0$, so the $C_\ell$
model collapses to a one-parameter fit of the total anisotropic power.

Four statistics are in play, and it helps to name them once. The first is
the full point-source matched filter of Sec.~\ref{sec:sd_coherent}, which
assumes a known source direction. The second is the unknown-direction
coherent scan, which maximizes that filter over the sky. The third is the
$C_\ell$ statistic, the total anisotropic power carried by the block powers
$C_L^P$. The fourth is the source-subspace covariance test, which fits the
amplitude of a source-shaped component of the pair covariance without
fitting its phase.

The third and the fourth are easily conflated, so we separate them before
comparing. Profiling the amplitude of the source-subspace covariance gives
a likelihood ratio that is a monotone function of the coherent statistic.
Two monotone statistics have the same receiver operating characteristic, so
once the thresholds are calibrated the covariance test and the coherent
search rank detections identically. What costs sensitivity is therefore not
fitting a covariance rather than a mean. It is averaging over the
$m$-pattern, which is what the $C_\ell$ compression does.

The fair comparison, when the source direction is unknown, is between the
scan and the $C_\ell$ statistic at the same angular resolution $L$. A
search of resolution $L$ has $K_L = L(L+2)$ anisotropy modes. The scan
maximizes the matched filter over the sky and pays a trials penalty of
about $K_L$ independent beams. The $C_\ell$ statistic uses the one number
those modes carry after the phases are averaged out, and is distributed as
$\chi^2_{K_L}$ under the null.

The two respond to the source differently, and that is the crux of the
comparison. The scan responds linearly to the source amplitude, so its
significance is $\rho_{\rm ps}$ itself, degraded only by the trials factor.
The $C_\ell$ statistic responds quadratically, because it is built from
squares of the $a_{Lm}$ and there is no phase left for the null to cancel
against. Its effective significance is $\rho_{\rm ps}^2/\sqrt{2K_L}$, which
is small whenever $\rho_{\rm ps}\lesssim1$, the regime the current data are
in. The threshold calculation below shows that the $C_\ell$ statistic is
never the more sensitive of the two at any amplitude. At the low angular
resolution PTAs have, however, it is not far behind.

\begin{figure*}
\centering
\includegraphics[width=\textwidth]{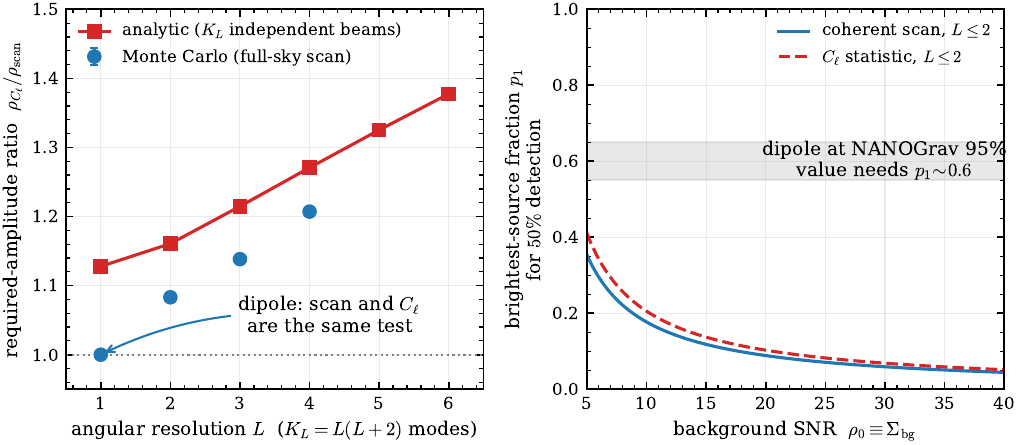}
\caption{\emph{Left:} the source amplitude required to detect via the
one-parameter $C_\ell$ statistic, relative to an unknown-direction
coherent scan, versus angular resolution $L$. The analytic non-central-$\chi^2$
estimate (squares), which treats the scan as $K_L = L(L+2)$ independent beams,
under-estimates the true scan threshold (the continuum maximum over directions
exceeds the maximum of $K_L$ independent beams) and so bounds the penalty from
above. The full-sky-scan Monte Carlo (circles) gives the smaller true penalty.
At $L=1$ the two searches are identical (true ratio $1.000$).%
\dataref{cl_penalty_mc_ratio_L1}{1.000}
\emph{Right:} the brightest-source
fraction $p_1$ needed for a $50\%$ detection at $p_{\rm fa}=0.05$, versus the
background SNR $\rho_0$, for the scan and the $C_\ell$ statistic at resolution
$L\le2$ (independent-beam thresholds). The shaded band marks $p_1\sim0.6$, the
fraction a dipole at the published NANOGrav $95\%$ value would require
(Sec.~\ref{sec:dipole_source}). The Monte Carlo error bars are the standard
errors of eight batch threshold estimates with common random variates:
$\pm0.005$ at $L=2$ and $L=3$, $\pm0.004$ at $L=4$, and exactly zero at
$L=1$, where the identity is enforced in the code and in the unit tests.%
\dataref{cl_penalty_mc_ratio_se_L2}{0.005}%
\dataref{cl_penalty_mc_ratio_se_L3}{0.005}%
\dataref{cl_penalty_mc_ratio_se_L4}{0.004}
\genby{scripts/fig\_cl\_vs\_scan\_penalty.py}}
\label{fig:cl_penalty}
\end{figure*}

Figure~\ref{fig:cl_penalty} (left) gives the source amplitude the $C_\ell$
statistic requires, relative to the coherent scan. The two searches are
exactly equal at $L=1$,\dataref{cl_penalty_mc_ratio_L1}{1.000} where the
scan maximized over directions returns the dipole-vector norm, which is the
number the $C_\ell$ statistic already uses. Beyond the dipole the $C_\ell$
statistic needs $8.3\pm0.5\%$ more amplitude at $L=2$,%
\dataref{cl_penalty_mc_percent_L2}{8.3}\dataref{cl_penalty_mc_percent_se_L2}{0.5}
$13.8\pm0.5\%$ at $L=3$,%
\dataref{cl_penalty_mc_percent_L3}{13.8}\dataref{cl_penalty_mc_percent_se_L3}{0.5}
and $20.7\pm0.4\%$ at $L=4$,%
\dataref{cl_penalty_mc_percent_L4}{20.7}\dataref{cl_penalty_mc_percent_se_L4}{0.4}
in the full-sky-scan Monte Carlo of the figure. An analytic
independent-beam estimate bounds the penalty from above at $38\%$ by
$L=6$.\dataref{cl_penalty_analytic_L6}{1.378} The right panel converts this
into the brightest-source fraction needed for a detection as a function of
$\rho_0$. The $C_\ell$ statistic is never stronger than the scan. It is
exactly as strong at $L=1$ and weaker for $L\ge2$.

That modest detection penalty coexists with a much steeper estimation
penalty, and the estimation penalty is the sharper statement of what
$C_\ell$ costs. The Fisher information the $C_\ell$ statistic carries about
the source fraction is itself proportional to $p_1^2$, so its
Kullback--Leibler divergence from the isotropic null scales as $p_1^4$. The
two source searches sit above it and scale differently from each other. The
pair-covariance matched filter has $\rho_{\rm ps}\propto p_1$
(Eq.~\ref{eq:rho_ps}), hence a divergence $\propto p_1^2$. A
continuous-wave search in the timing-residual mean fits a strain amplitude
rather than a power, and that amplitude scales as $\sqrt{p_1}$, so its
divergence scales as $p_1$. The error bar on $p_1$ from $C_\ell$ alone is
worse than the coherent one by a factor $\propto1/(p_1\Sigma_{\rm bg})$,
where $\Sigma_{\rm bg} = \rho_0$ is the background significance of
Eq.~\ref{eq:sigma_bg}, and that factor diverges for faint sources. A
low-resolution $C_\ell$ statistic can thus detect a source at a modest
amplitude penalty while characterizing it far more poorly. For detection
and for characterization alike, the $C_\ell$ statistic never does better
than the source search.

\subsection{Implications}
\label{sec:sd_bottom}

Two conclusions follow, one for the present and one for the future.

At present there is no anisotropy measurement to be had. With a
per-bin cross-correlation background SNR of $\approx 2$
(Sec.~\ref{sec:sd_setup}), even the optimal coherent search of a source
carrying $p_1\sim0.1$ reaches only $\rho_{\rm ps}\lesssim1$, and the
$C_\ell$ statistic, weaker and quadratic in the source amplitude,
is farther below threshold still. The most powerful search of all is the
coherent one, in the limiting case a continuous-wave search in the
timing-residual mean. It has been carried out on the NANOGrav 15-year
data by Goncharov et al.~\cite{Goncharov:2026joint}, who find no
resolvable source and infer a number of contributing sources consistent
with the Gaussian-background limit. (Their public analysis uses the
earlier, partly simulated version of the 15-year dataset that an
inter-array data-sharing rule permits showing. The authors report that
the search on the final 15-year data gives an essentially identical
result.) They forecast the probability of detecting an individual source
at signal-to-noise above $5$ to be only $\sim\!2\%$ in the current
15-year data, rising to $\sim\!5\%$ in the projected 20-year array. If
the optimal search cannot find the source, no $C_\ell$ statistic can
find its anisotropy: the same source powers both.

The future is a question of how well the background is measured. If the
per-bin SNR eventually reaches $\rho_0\simeq20$, the right panel of
Fig.~\ref{fig:cl_penalty} gives the source fractions a search would then
reach. A coherent scan at quadrupole
resolution detects a source carrying $p_1\gtrsim9\%$ of the power, a
dipole-only search needs $p_1\gtrsim16\%$, and the full point-source
matched filter reaches $p_1\gtrsim7\%$ (scanning the full template over
$K_L = 48$ beams at $L=6$, where the template has converged).%
\dataref{required_p1_rho020_scanL2}{0.089}%
\dataref{required_p1_rho020_dipole}{0.164}%
\dataref{required_p1_rho020_fullps}{0.069}
The
$C_\ell$ statistic at quadrupole resolution then needs
$8.3\pm0.5\%$ more%
\dataref{cl_penalty_mc_percent_L2}{8.3}%
\dataref{cl_penalty_mc_percent_se_L2}{0.5}
source amplitude than the scan (Fig.~\ref{fig:cl_penalty}, left). The
more sensitive way to search, then and now, is for the source itself,
and at current sensitivity neither statistic has anything to find.
\section{The published \texorpdfstring{$C_\ell$}{Cl} upper limits and the analysis prior}
\label{sec:nanograv}

The previous two sections leave a puzzle. The NANOGrav 15-year
anisotropy analysis~\cite{NANOGrav:2023tcn} quotes 95\% upper limits on
$C_\ell/C_0$: at the dipole, $C_1/C_0 \lesssim 0.2$. But we have just
seen that the data cannot at present inform any anisotropy statistic.
The anisotropy is one bright source, that source is below current
detection thresholds even for the optimal search
(Secs.~\ref{sec:dipole_source}--\ref{sec:source_detection}), and the
$C_\ell$ statistic is weaker than the optimal search. A Bayesian
analysis applied to data that carry no information about a parameter
returns the prior on that parameter, so we expect the published bound to
reproduce the prior of the analysis basis. In this section we show that
it does.

\subsection{The square-root spherical harmonic basis}

The NANOGrav analysis uses the square-root spherical harmonic
basis~\cite{Banagiri:2021lisa}, in which the GW power is written as
\begin{equation}
P(\hat{\Omega}) = \left[\sum_{L=0}^{L_{\rm max}} \sum_{M=-L}^{L}
b_{LM} Y_{LM}(\hat{\Omega})\right]^2,
\end{equation}
ensuring positivity. Here $P(\hat\Omega)$ is the GW power distribution
on the sky, the same quantity denoted $M(\hat n)$ in
Sec.~\ref{sec:cl}. The priors on the coefficients are: $b_{00} = 1$
(fixed), $|b_{LM}| \sim U[0, 50]$ with uniform phase for $M \neq 0$,
and $b_{L0} \sim U[-50, 50]$. The ratio $C_\ell/C_0$ is invariant under
a common rescaling of the $b_{LM}$, and the prior range ($50$) is large
compared with the fixed $b_{00} = 1$, so the prior induced on
$C_\ell/C_0$ is essentially independent of the amplitude range. It is
set by the basis and its truncation.

The analysis uses $\ell_{\rm max}^a = 6$ for the measured $C_\ell$.
Following Banagiri et al.~\cite{Banagiri:2021lisa}, the $b_{LM}$
expansion is truncated at $\ell_{\rm max}^b = \ell_{\rm max}^a / 2 = 3$.
This follows from the Clebsch--Gordan selection rule: squaring a field
with modes up to $L$ produces power up to $\ell = 2L$.

\subsection{The prior and the published upper limit}

We sample $5\times10^4$ realizations from the $b_{LM}$ prior with
$\ell_{\rm max}^b = 3$ and compute $C_\ell/C_0$ by squaring the
resulting field. The 95th percentile of $C_1/C_0$ under this prior is
$0.200$,\dataref{sqrtSH_prior_pct95}{0.2005} matching the broadband
95\% upper limit quoted by NANOGrav,
$C_{\ell>0}/C_0 < 20\%$~\cite{NANOGrav:2023tcn}. NANOGrav's own
Hellinger-distance analysis reaches the same conclusion: in each of the
lowest five frequency bins the posterior on $C_\ell/C_0$ is
statistically indistinguishable from the prior~\cite{NANOGrav:2023tcn}.
This is not a criticism of the analysis. It is the expected outcome
given Sec.~\ref{sec:source_detection}. The data had nothing to say about
anisotropy, and the machinery correctly returned the assumption it was
handed. The point is about interpretation: the number $0.2$ reflects the
basis and its truncation rather than the sky, and model comparisons that
treat it as a measurement
(e.g.~\cite{LinLidzMa:2026,LinLidzMa:2026b}) are comparing to
the prior rather than to the data.

Table~\ref{tab:percentiles} compares the percentiles of the sqrt-SH
prior with the three astrophysical models at the three representative
frequencies. At the lowest frequency, where the PTA is most sensitive,
even the most anisotropic model ($\varepsilon = 0.66$) has a median
$C_1/C_0 = 0.050$, close to the prior's median of $0.065$ but with a
completely different distributional shape. The low-scatter models sit
orders of magnitude below.%
\dataref{c1c0_median_eps066_f0085}{0.050}\dataref{sqrtSH_prior_pct50}{0.0654}

\begin{table}
\centering
\caption{Percentiles of $C_1/C_0$: the sqrt-SH prior
($\ell_{\rm max}^b = 3$, frequency-independent, $5\times10^4$ samples)
compared with three astrophysical models at three frequencies. The
prior's 95th percentile matches the published NANOGrav 95\% upper
limit. \genby{scripts/table\_percentiles.py}}
\label{tab:percentiles}
\begin{tabular}{l c c c}
\hline\hline
& 5th & 50th & 95th \\
\hline
\multicolumn{4}{l}{\textit{sqrt-SH prior ($\ell_{\rm max}^b = 3$)}} \\
\quad (all frequencies) & 0.0087 & 0.0654 & 0.200 \\[4pt]
\multicolumn{4}{l}{\textit{$\varepsilon = 0.20$}} \\
\quad $f = 0.085$ yr$^{-1}$ & 0.00009 & 0.00078 & 0.0102 \\
\quad $f = 0.28$ yr$^{-1}$ & 0.00085 & 0.0079 & 0.100 \\
\quad $f = 1.03$ yr$^{-1}$ & 0.0054 & 0.0499 & 0.439 \\
\multicolumn{4}{l}{\textit{$\varepsilon = 0.38$}} \\
\quad $f = 0.085$ yr$^{-1}$ & 0.00051 & 0.0051 & 0.0872 \\
\quad $f = 0.28$ yr$^{-1}$ & 0.0026 & 0.0254 & 0.295 \\
\quad $f = 1.03$ yr$^{-1}$ & 0.0092 & 0.0897 & 0.634 \\
\multicolumn{4}{l}{\textit{$\varepsilon = 0.66$}} \\
\quad $f = 0.085$ yr$^{-1}$ & 0.0046 & 0.0505 & 0.537 \\
\quad $f = 0.28$ yr$^{-1}$ & 0.0097 & 0.104 & 0.693 \\
\quad $f = 1.03$ yr$^{-1}$ & 0.0200 & 0.191 & 0.849 \\
\hline\hline
\end{tabular}
\end{table}

\begin{figure*}
\centering
\includegraphics[width=\textwidth]{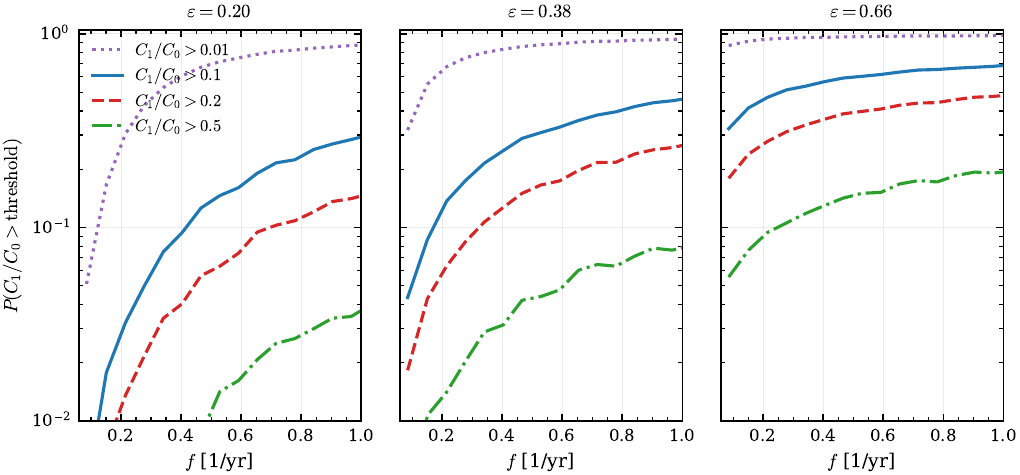}
\caption{Probability that $C_1/C_0$ exceeds a given threshold as a
function of GW frequency, for three astrophysical models. The
$C_1/C_0 > 0.2$ curve corresponds to the published NANOGrav 95\%
upper limit, which coincides with the prior percentile (see text). The
other thresholds show what limits of different depths would
discriminate. Even for the most anisotropic model
($\varepsilon = 0.66$), the $0.2$ threshold is
exceeded in only $\sim 18\%$ of realizations at the lowest frequency.
The probability rises at high frequency (where fewer sources contribute),
but this is also where PTA sensitivity is weakest.
\genby{scripts/fig\_exceedance\_probability.py}}
\label{fig:exceedance}
\end{figure*}

Figure~\ref{fig:exceedance} shows the probability of exceeding fixed
$C_1/C_0$ thresholds as a function of frequency for the three
astrophysical models. Even for the most anisotropic model
($\varepsilon = 0.66$), the probability that $C_1/C_0 > 0.2$ is only
$\sim 18\%$ at the lowest frequency, rising to $\sim 48\%$ at
$f = 1.03\,{\rm yr}^{-1}$. The high-frequency end, however, is precisely
where the PTA loses anisotropy sensitivity. For the $\varepsilon = 0.20$
model the threshold is essentially never reached at the sensitive low
frequencies ($P = 0.2\%$ at $f = 0.085\,{\rm yr}^{-1}$), climbing to
$\sim 15\%$ only at $f = 1.03\,{\rm yr}^{-1}$. The other thresholds in
the figure bracket the same conclusion. A threshold of $0.01$ is
exceeded by nearly every realization of every model except
$\varepsilon = 0.20$ at the lowest frequencies, so a limit at that level
would begin to bite. A threshold of $0.5$ is rare even for
$\varepsilon = 0.66$.
A limit at the level of the quoted $0.2$, even if it were
data-driven, would therefore not usefully discriminate among the models
at the frequencies where the PTA is sensitive.

\subsection{Sensitivity to $\ell_{\rm max}^b$}

The prior distribution of $C_\ell/C_0$ depends on
$\ell_{\rm max}^b$. For the NANOGrav choice $\ell_{\rm max}^b = 3$,
the 95th percentile of $C_1/C_0$ is $\approx 0.20$. It drops rapidly
with $\ell_{\rm max}^b$: $0.086$ (6), $0.0088$ (23), $0.0022$ (47),
$0.00056$ (95), computed by sampling the same prior at each
truncation.\dataref{sqrtSH_prior_pct95_vs_lmaxb}{0.086/0.0088/0.0022/0.00056} This is because the squaring operation couples all
$b_{LM}$ modes: adding more high-$L$ modes pumps power into $C_0$
through the convolution, shrinking $C_\ell/C_0$. A future analysis that
merely raised the truncation would therefore report a much tighter
limit from the same uninformative data. Beyond the overall level, the
shape of the prior distribution at any $\ell_{\rm max}^b$ bears
no resemblance to the astrophysical distribution from discrete SMBHB
populations (Sec.~\ref{sec:dipole_source}).
\provnote{The $\ell_{\rm max}^b$ sequence is computed by
\texttt{scripts/mc/run_sqrtSH_realizations.py} run at the four
truncations.}
\section{Discussion}
\label{sec:discussion}

\subsection{What constrains the astrophysical model now}

The logic of the preceding sections assembles into a simple ranking of
observables, ordered by how much they can teach us about the SMBHB
population with current and near-term data.

\emph{The shape of the strain spectrum.} This is the most informative
observable today, and the only one already delivering constraints. The
heavy-tailed, high-$M_{\rm peak}$ models that maximize the anisotropy
are the same ones that depress the median spectrum below its $f^{-4/3}$
mean and inflate its frequency-to-frequency scatter
(Sec.~\ref{sec:mean_median}). At $\varepsilon = 0.66$ the median is
$46\%$ of the mean at the lowest frequency and $5\%$ at $f =
1.03\,{\rm yr}^{-1}$. Fitting the NANOGrav 15-year free spectrum to the
full predicted distribution $p(h_c^2(f)\mid\vec\theta)$ already requires
roughly ten times more black holes than local scaling relations suggest
and disfavors $M_{\rm peak}\gtrsim10^{10}\,M_\odot$ at
$2\sigma$~\cite{SatoPolito:2025dist}. That fit uses purely spectral,
one-dimensional information: no inter-pulsar correlation measurement,
let alone an anisotropy measurement, is involved.

\emph{A direct search for the brightest source.} This is the right way
to look for the anisotropy, but it is some distance behind. The
significance of the optimal search is $\rho_{\rm ps} =
\sqrt5\,p_1\,\rho_0$ (Sec.~\ref{sec:source_detection}), and with the
Hellings--Downs cross-correlation detected at a total SNR of
$\approx5$ over the entire band~\cite{NANOGrav:2023gor}, which is
$\rho_0\approx2$ per frequency bin, no astrophysically plausible source
is within reach. The joint resolved-plus-unresolved search of the
15-year data confirms this empirically: no resolvable source, with a
$\sim\!2\%$ ($15$ yr) to $\sim\!5\%$ ($20$ yr) forecast probability of
an ${\rm SNR}>5$ detection~\cite{Goncharov:2026joint}. As the
background cross-correlation SNR grows, this channel will open. At
$\rho_0\simeq20$ per bin, a quadrupole-resolution scan detects sources
carrying $p_1\gtrsim9\%$ of the power (Fig.~\ref{fig:cl_penalty}), and
the full point-source matched filter reaches $p_1\gtrsim7\%$ under the
trials assumption of Sec.~\ref{sec:sd_bottom}.

\emph{The angular power spectrum.} Never ahead of the source search,
equal to it at $L=1$, and behind for $L\ge2$. The reason is that
$C_\ell$ discards the phase information that locates the source, at a
modest cost in detection threshold at low resolution and a severe cost
in characterization (Sec.~\ref{sec:sd_cl}). It is also uninformative at
present: the published bounds reproduce the prior of the analysis
basis (Sec.~\ref{sec:nanograv}). Since the same brightest source powers
both statistics, a $C_\ell$ detection before a source detection should
not be expected, and a $C_\ell$ bound obtained in the absence of a
source detection should be examined for prior content.

\subsection{Relation to recent work}

Lin, Lidz \& Ma~\cite{LinLidzMa:2026} compute the expected shot-noise
$C_\ell$ from empirically calibrated merger models and compare with
the NANOGrav bounds, concluding that near-future PTA data should be
capable of detecting this signal or placing meaningful constraints on
SMBHB merger models.

While the present work was being completed, the same authors published
Monte Carlo realizations of that signal~\cite{LinLidzMa:2026b}. That
paper revises the earlier calculation in four places. Its moment-based
estimate exceeds its own Monte Carlo mean by a factor of $130$ at
$f = 1\,{\rm yr}^{-1}$. The $f^{8/3}$ frequency scaling of the earlier
paper is replaced by $f^{1.1}$ to $f^{1.2}$. The per-realization shot
noise is shown to satisfy $\hat C_{\rm shot}/4\pi \le 1$. And the
effective number of sources is taken to be the realized
$1/\sum_a p_a^2$. Our Sec.~\ref{sec:astro} finds the same separation
between the ensemble mean and the typical realization, in the same
direction and for the same reason. The definition of $N_{\rm eff}$ is
the one used here (Eq.~\ref{eq:neff}), and their bound is
$N_{\rm eff}\ge1$ written in the other notation.

Two of our qualifications are addressed by neither paper. Both papers
compare against NANOGrav bounds that reproduce the prior of the analysis
basis rather than a measurement (Sec.~\ref{sec:nanograv}), so a predicted
median a factor of four below that bound is not a factor of four from
being measured. And when anisotropy does become measurable, the same
information will be available first, and more powerfully, from the
direct source search (Sec.~\ref{sec:source_detection}).

\subsection{Summary}

The gravitational wave background from SMBH binaries is anisotropic
because it is sourced by a finite number of discrete objects. For every
population model consistent with the measured amplitude, essentially
all of that anisotropy is produced by one bright binary. A dipole at
the published NANOGrav 95\% value would require $p_1 \sim 0.6$
(Sec.~\ref{sec:dipole_source}). That binary is currently undetectable.
Its strain lies below the sky-averaged continuous-wave limit at every
frequency that limit covers, the optimal search of the 15-year data
finds nothing, and
the odds of a near-term detection are at the percent level. Because
every anisotropy statistic is at best as sensitive as the direct source
search, no anisotropy measurement is possible with current data, and
the published $C_\ell/C_0$ upper limits reproduce the prior of the
square-root spherical-harmonic basis rather than a property of the sky.
What the data do constrain, through the shape and scatter of the strain
spectrum with no angular information at all, is precisely the property
that controls the future anisotropy: how heavy the brightest sources
are. By the time the background cross-correlation is measured well
enough for anisotropy to be within reach, the spectrum will have
narrowed the candidate models further, and the first anisotropy
detection is more likely to arrive as a source detection than as a
$C_\ell$ measurement.
 
\begin{acknowledgments}
This work was done entirely using Claude Code and Codex. All of the
code, data and \LaTeX{} source behind this paper are publicly
available at \url{\repourl}, and the production Monte Carlo arrays are
archived at \url{\zenodourl}. GSP acknowledges support from the
Friends of the Institute for Advanced Study Fund. MZ
acknowledges support from the National Science Foundation NSF-BSF
2207583 and NSF 2209991, the Nelson Center for Collaborative Research
and the Simons Foundation through the Black Holes and Strong Gravity
program through Award No.\ SFI-MPS-BH-00012593-10.
\end{acknowledgments}

\bibliography{references}

\end{document}